\documentclass[aps,prb,twocolumn,showpacs,showkeys,superscriptaddress]{revtex4-1}
\usepackage{graphicx}
\begin{document}

\title{Cyclic electric field stress on bipolar resistive switching devices}

\author{A. Schulman}
\thanks{University of Buenos Aires and CONICET of Argentina scholarships}
\affiliation{Laboratorio de Bajas Temperaturas - Departamento de
F\'{\i}sica - FCEyN - Universidad de Buenos Aires and IFIBA -
CONICET,  Pabell\'on I, Ciudad Universitaria, C1428EHA Buenos Aires,
Argentina}
\author{C. Acha}
\thanks{corresponding author (acha@df.uba.ar)}
\address{Laboratorio de Bajas
Temperaturas - Departamento de F\'{\i}sica - FCEyN - Universidad de
Buenos Aires and IFIBA - CONICET,  Pabell\'on I, Ciudad
Universitaria, C1428EHA Buenos Aires, Argentina}

\date{\today}


\begin{abstract}

We have studied the effects of accumulating cyclic electrical pulses
of increasing amplitude on the non-volatile resistance state of
interfaces made by sputtering a metal (Au, Pt) on top of the surface
of a cuprate superconductor YBa$_2$Cu$_3$O$_{7-\delta}$ (YBCO). We
have analyzed the influence of the number of applied pulses $N$ on
the relative amplitude of the remnant resistance change between the
high ($R_H$) and the low ($R_L$) state
[($\alpha=(R_{H}-R_{L})/R_{L}$] at different temperatures ($T$). We
show that the critical voltage ($V_c$) needed to produce a resistive
switching (RS, i.e. $\alpha >0$) decreases with increasing $N$ or
$T$. We also find a power law relation between the voltage of the
pulses and the number of pulses $N_{\alpha_0}$ required to produce a
RS of $\alpha=\alpha_0$. This relation remains very similar to the
Basquin equation used to describe the stress-fatigue lifetime curves
in mechanical tests. This points out to the similarity between the
physics of the RS, associated with the diffusion of oxygen vacancies
induced by electrical pulses, and the propagation of defects in
materials subjected to repeated mechanical stress.

\end{abstract}



\maketitle

\section{INTRODUCTION}

The search of new non-volatile memories is reinforced nowadays by
the necessity of producing more dense, less dissipative and low cost
devices.~\cite{Burr08} Memories based on the resistive switching
(RS) mechanism on metal-oxide interface are marked as one of the
most promising candidates for the next generation of memory
applications.~\cite{Waser07,Sawa08} A typical RS device consists of
an interface between a metal and an oxide which can be either in a
capacitor-like form or in a planar structure. Depending on the
oxide, the mechanism beneath the RS can give rise to a
polarity-independent filamentary effect (typically observed for
binary oxides)~\cite{Kim11} or to a polarity-sensitive one (observed
for complex-oxides) associated with interfacial properties
~\cite{Jeong12,Schindler07}, although exceptions can be observed in
both categories.~\cite{Fujimoto06,Li11}

While significant advances have been made in improving device
performances, understanding their underlying physics still
represents a great challenge. In addition to scalability, fast
response, repeatability, retentivity and low power consumption,
endurance is one of the properties that these memories must
fulfill.~\cite{Yang13} The cyclic electric field stress that implies
the repeated switching of the device may produce an accumulation of
defects that would affect its electrical properties. As during the
normal operation of a device out of the range high and low
resistance states can be produced, error correction techniques must
be taken into account.~\cite{Stievano11} These techniques can be
based on feedback protocols in order to achieve, for example, a RS
with resistance values in the desired range.

This is the particular point that we address in this paper, by
analyzing the evolution of the remanent resistance of a bipolar
device composed by metal-perovskite [(Au,Pt)-YBCO] junctions upon
the application of a cyclic accumulation of pulses. In this type of
devices, by considering that the resistance of the interface is
proportional to the density of vacancies, the RS mechanism was
associated with the electromigration of oxygen
vacancies.~\cite{Rozenberg10} Our results indicate that the
mechanism that determines the evolution of the remnant resistance
upon the application of a cyclic accumulation of pulses presents
close similarities to the one governing the propagation of fractures
during a mechanical-fatigue test on a material.\cite{Suresh98}

\section{EXPERIMENTAL}

To study the dependence of the bipolar RS with cyclic electric field
stress we sputtered four metallic electrodes on one of the faces of
a good quality YBCO textured ceramic sample (see the inset of
Fig.~\ref{fig:1}). The width of the sputtered electrodes was in the
order of 1 mm with a mean separation between them of 0.4 mm to 0.8
mm. They cover the entire width of one of the faces of the YBCO slab
(8x4x0.5 mm$^3$). Silver paint was used to fix copper leads
carefully without contacting directly the surface of the sample.
Details about the synthesis and the RS characteristics of the
metal-YBCO interfaces can be found
elsewhere.~\cite{Rozenberg10,Acha09a,Acha09b,Placenik10,Acha11,Placenik12,Schulman12}
We choose as metals Au and Pt for the pair of pulsed electrodes,
labeled $1$ and $2$, respectively. As we have shown previously, the
Pt-YBCO interfaces have a lower resistance value than the Au-YBCO
ones ($R(Pt)\lesssim R(Au)/3$), and a small RS amplitude. In this
way, only the Au-YBCO ($1$) electrode will be active, simplifying
the effects produced upon voltage pulsing treatments. After applying
a burst of $N$ ($10^4 \leq N \leq 5.10^5$) unipolar square voltage
pulses (100 $\mu$s width at 1 kHz rate), the remnant resistance of
each pulsed contact was measured using a small current through
contacts 1-2 and a convenient set of additional Au contacts. To
estimate the resistance $R(Au)$ and $R(Pt)$, of the active Au-YBCO
and of the Pt-YBCO interfaces, respectively, the voltage difference
between electrodes 1-3 and 4-2 was measured. Corrections to $R(Au)$
and $R(Pt)$ by considering the resistance of the bulk YBCO are
negligible taking into account that its value is only $\simeq 0.1
\Omega \ll 50 \Omega < R(Au),R(Pt)$ in this temperature range. We
want to note that the polarity of the pulses was defined arbitrarily
with the ground terminal connected to the Au-YBCO contact.
Temperature was measured with a Pt thermometer in the 200 K to 340 K
range and stabilized better than at 0.5\% after each pulsing
treatment.

To perform a cyclic electric field stress experiment at a fixed
temperature $T_0$, temperature is initially stabilized. As no
electroforming step is needed, we initially set the active Au-YBCO
electrode to its low resistance state ($R(Au)_L$) with a burst of
pulses of -5 V amplitude while, in a complementary manner, the
Pt-YBCO electrode is in its high resistance state ($R(Pt)_H$). Note
here that as the Au-YBCO electrode is the ground terminal a negative
pulse indicates that its potential is higher than that of the
Pt-YBCO electrode. Consequently the density of oxygen vacancies near
the Au-YBCO interface should decrease (as they are positive charged
defects), reducing the interfacial resistance, as we observe, in
accordance to the voltage enhanced oxygen vacancy model that
describes RS for bipolar devices.~\cite{Rozenberg10}

As mentioned previously, no relevant changes are expected in this
electrode resistance ($R(Pt)_L \simeq R(Pt)_H$). Then we apply a
"reset" burst of $N$ unipolar pulses with a $V_{pulse}$ amplitude
during a time $t_0$ (from 10 s to 500 s, depending on the $N$
value). Although the temperature is constantly stabilized to $T_0$,
in order to avoid overheating effects on the resistance
measurements, a time $t_0$ is waited before measuring the remnant
resistance of each pulsed contact ($R(Au)_H$ and $R(Pt)_L$). After
that, a "set" burst of maximum opposite polarity (-$V_{pulse}^{max}
= -5 V$) is applied to subject the material to a cyclic stress. In
this way the resistance change is partially recovered and both
remnant resistances are measured again ($R(Au)_L$ and $R(Pt)_H$).
The process is then completely repeated for a new $V_{pulse}$ value,
increased with a fixed step, until it reaches our experimental
maximum ($V_{pulse}^{max} = 5 V$).

\section{RESULTS AND DISCUSSION}

\begin{figure}
\vspace{7mm}
\centerline{\includegraphics[angle=0,scale=0.4]{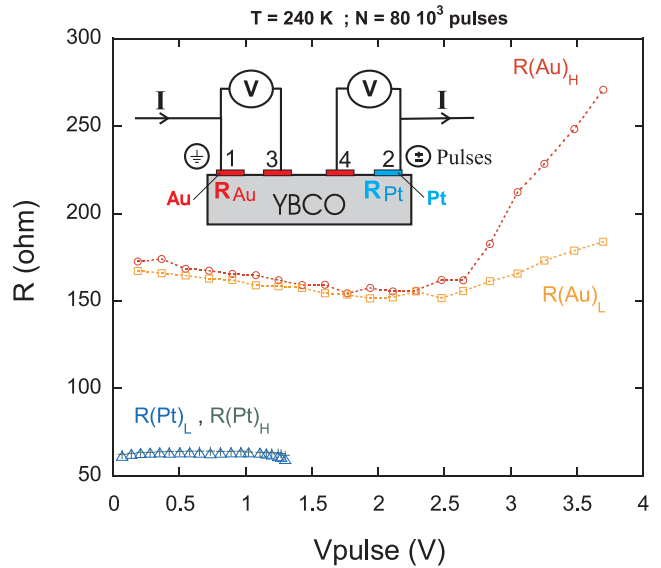}}
\vspace{-0mm}\caption{(Color online) Dependence with the amplitude
of $N$=80.$10^3$ square pulses ($V_{pulse}$) of the remanent
resistance of each contact (low state and high state for the Au and
Pt interfaces, respectively). The $V_{pulse}$ magnitude is different
for each contact as it corresponds to their effective voltage drop.
100 $\mu$s pulses were applied at 1 kHz rate at a constant
temperature (240 K). The high and the low state of these contacts
obtained after applying -$V_{pulse}^{max}$ is also shown for
comparison. Lines are guides to the eye.} \vspace{-0mm}
\label{fig:1}
\end{figure}

A typical result of a cyclic electric field stress experiment at a
fixed number of pulses of the burst ($N$ = 80 k pulses) and
temperature (240 K) is shown in Fig.~\ref{fig:1}, where we have
plotted the remnant resistances of each pulsed electrode (Au and Pt)
in both high and low states as a function of the "reset" amplitude
of the pulse ($V_{pulse}$). $V_{pulse}$ corresponds to the voltage
drop measured at each particular electrode. We can observe that, as
expected, $R(Pt)_H \simeq R(Pt)_L$ for the smaller voltage drop
explored, while $R(Au)_H > R(Au)_L$ for $V_{pulse}$ higher than a
critical voltage ($V_c$), indicating the existence of RS for this
electrode. Hereafter we will show results only for this Au-YBCO
active electrode.

The relative amplitude of the remnant resistance change between the
high ($R_H$) and the low ($R_L$) state is defined as
$\alpha=(R_{H}-R_{L})/R_{L}=\Delta R / R_{L}$. It can be noted that
while $\alpha \simeq 0$ for $V_{pulse} \leq V_c$ both $R(Au)_H$ and
$R(Au)_L$ decrease with increasing $V_{pulse}$. This is a
consequence of the protocol used for this particular cyclic
treatment that forces $R(Au)_L$ to a lower value than the initial
one, as for each reset burst of amplitude $V_{pulse}$ a set burst of
amplitude -$V_{pulse}^{max}$ is applied. This situation is reversed
for $V_{pulse} \geq V_c$ were the increase in $R(Au)_H$ obtained is
not completely recovered when the set protocol is applied.

\begin{figure}
\vspace{7mm}
\centerline{\includegraphics[angle=0,scale=0.4]{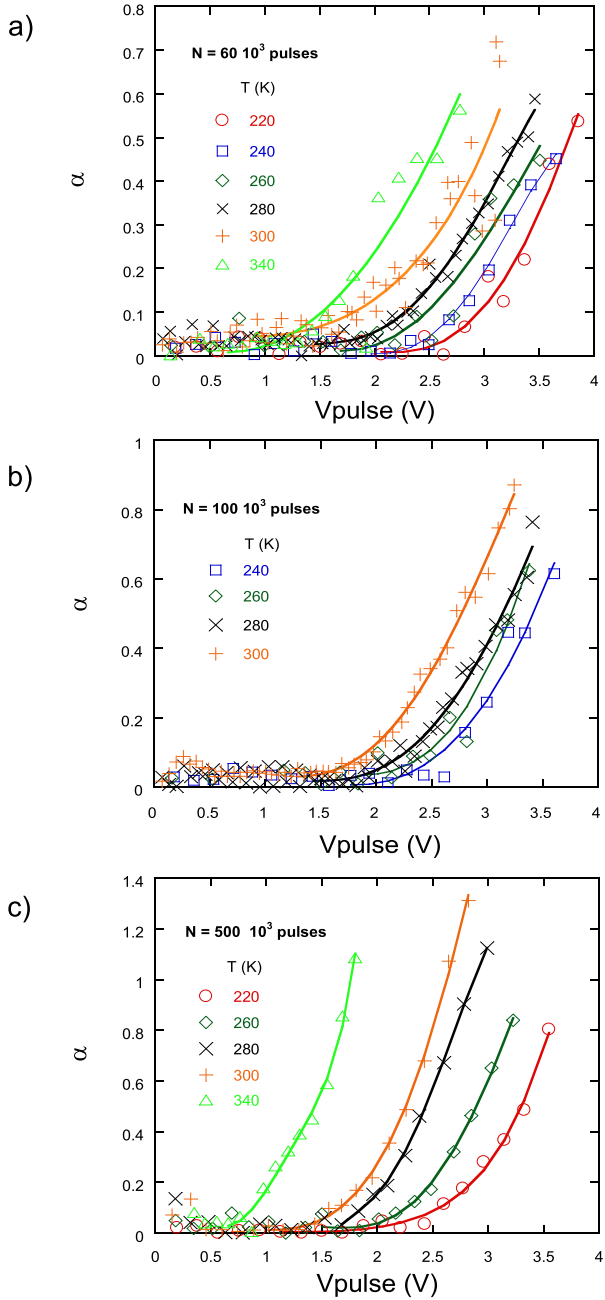}}
\vspace{-0mm}\caption{(Color online) Relative variation of the
remanent resistance ($\alpha=\Delta R/R_{L}$) of the Au-YBCO
interface as a function of the amplitude of the reset pulses
($V_{pulse}$) at different temperatures. The pulse treatment
corresponds to a) N = 60 10$^3$ b) N = 100 10$^3$ and c) N = 500
10$^3$ square pulses with the same characteristics described in the
text. Lines are guides to the eye.} \vspace{-0mm} \label{fig:2}
\end{figure}

\begin{figure}
\vspace{7mm}
\centerline{\includegraphics[angle=0,scale=0.4]{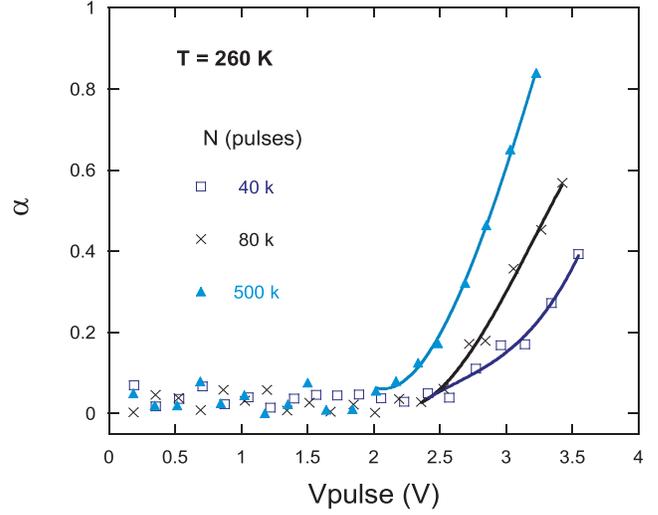}}
\vspace{-0mm} \caption{(Color online) $\alpha$ as a function of the
amplitude of the reset pulses ($V_{pulse}$) at 260 K, varying the
number $N$ of applied pulses. Lines are guides to the eye.}
\vspace{-0mm} \label{fig:3}
\end{figure}

The dependence of $\alpha$ with the amplitude $V_{pulse}$ of the
burst of $N$ pulses at different temperatures can be depicted in
Fig.~\ref{fig:2}. A noisy $\alpha < 0.1$ is obtained until
$V_{pulse} \geq V_c$, where $\alpha$ increases with
($V_{pulse}-V_c$), following a power law-like behavior. As
temperature is increased, $V_c$ decreases linearly and $\alpha$
reaches higher values. A similar behavior occurs when performing the
experiment at a fixed temperature and increasing the number $N$ of
pulses conforming the burst, as it is shown in Fig.~\ref{fig:3}. In
this case, $\alpha$ and $V_c$ varies logarithmically with $N$ (not
shown here). The former was also observed for Ag-manganite
interfaces when $N$ pulses of the same polarity were accumulated.~
\cite{Ghenzi12}

In a mechanical fatigue test the material is subjected to a cyclic
stress. Fatigue occurs when, for a stress ($S$) above a certain
threshold, a defect zone progresses along the material until a
fracture occurs for $N_\alpha$ cycles, which corresponds to the
number of cycles to failure. $S-N_\alpha$ curves (or W\"{o}hler's
curves\cite{Schutz96}) can then be plotted in order to represent the
lifetime of a meterial subjected to these conditions.

Here, for bipolar devices, it is known that the RS is associated
with the electro-migration of vacancies in and out of the active
metal-oxide interface.~\cite{Rozenberg10} We can expect that a
severe pulsing treatment may generate extended vacancy defects, but,
as far as we know, in the electric field range explored, no
macroscopic cracks were ever observed.~\cite{Moeckly93} So, it is
necessary to define what can we call as an electric failure of the
device. As possible scenarios, it can be expected that an arbitrary
value of $\alpha$ can be obtained if the density of vacancies near
the active interface (ai) can be reversibly increased and decreased
making that $R(ai)_H \gg R(ai)_L$. In this case there is not a
proper failure. Instead, more than a failure criteria, a value to
achieve can be established, as for example $\alpha \geq \alpha_0$,
which can be actually considered as a condition to fulfill for the
reliable operation of the device. This can be the particular case of
our devices, where a moderate RS can be obtained ($\alpha \lesssim
1$), with $\alpha$ increasing in the whole $V_{pulse}$ range
explored.

Additional scenarios can be considered, like the case where $\alpha$
saturates, reaching a maximum, indicating that a dynamic equilibrium
is established between the number of vacancies generated by the
pulses and the vacancies filled with oxygens. Other possible failure
scenario is the one that can be observed in unipolar devices, where
a failure occurs when the low state requires a very high current to
be reset\cite{Lee08}. In this case, a proper RS failure is obtained
and corresponds to $\alpha=0$ in the whole operating $V_{pulse}$
range.

\begin{figure}
\vspace{7mm}
\centerline{\includegraphics[angle=0,scale=0.4]{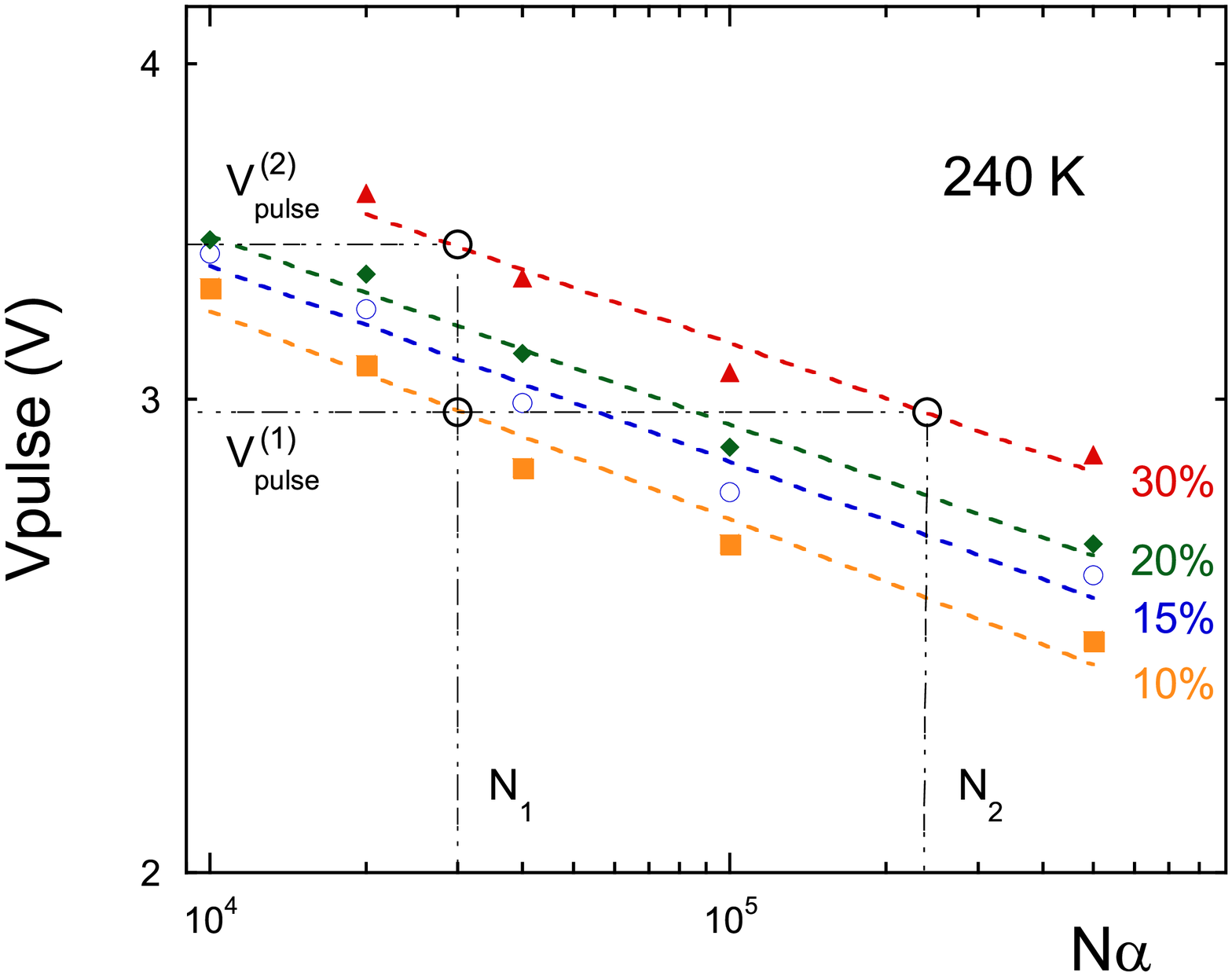}}
\vspace{-0mm}\caption{(Color online) Electric field stress lifetime
curves (V-N curves) at 240 K, where the failure criteria corresponds
to an arbitrarily value of $\alpha$ = 10 to 30 \%. Dotted lines are
fits corresponding to Eq.~\ref{eq:Basquin}. Slash-dotted lines
indicate two possible correction protocols to modify an obtained
$\alpha_1$=10\% to a targeted $\alpha_2$=30\% (see text).}
\vspace{-0mm} \label{fig:4}
\end{figure}

As mentioned previously, in our experimental case we choose to
define arbitrarily that a failure occurs when $\alpha$ reaches a
predetermined value ($\alpha_0$). With the results presented in Fig.
2 and 3, a typical stress-fatigue lifetime curve (or $V-N_\alpha$
curve), shown in Fig.~\ref{fig:4}, can be obtained as an answer to
the question of which is the number of pulses $N_{\alpha_0}$ of
amplitude $V_{pulse}$ needed to produce an arbitrarily fixed value
of $\alpha=\alpha_0$ at a temperature $T=T_0$ (240 K). Different
values of $\alpha_0$ were considered (10\%, 15\%, 20\%, 30\%) to
check the sensitivity to its particular value. Independently of the
value of $\alpha_0$, a power law is obtained between $V_{pulse}$ and
$N_{\alpha}$, which, in fact, can be associated with the Basquin
equation that describes the $S-N_\alpha$ curves in material fatigue
experiments:

\begin{equation}
\label{eq:Basquin} V_{pulse} = A N_{alpha}^{\beta},
\end{equation}

\noindent where $A$ is a constant and $\beta$ the Basquin exponent.

Independently of the failure criteria adopted we obtained very
similar curves with nearly the same exponent $-0.1 \lesssim \beta
\lesssim -0.07$. Surprisingly, this particular value is typical for
most of the mechanical fatigue test performed on
metals.\cite{Suresh98}

If after a reset (or a set) pulsing protocol of $N_1$ pulses of
amplitude $V_{pulse}^{(1)}$ the obtained $\alpha_1$ is out of the
targeted range ($\alpha_2 \pm \Delta\alpha$), the results showed in
Fig.~\ref{fig:4} can be very useful in order to establish a first
order feedback protocol to correct this issue with a single burst of
pulses. By considering the sensitivity to $\alpha$ of the Basquin
curves at a fixed $V_{pulse}^{(1)}$ or at a fixed number of pulses
$N_1$, two different strategies can be envisioned: to apply a new
burst of $N_2$ pulses at fixed $V_{pulse}^{(1)}$, or to fix the
number of pulses to $N_1$ and modify the amplitude $V_{pulse}^{(2)}$
(see Fig.~\ref{fig:4}). In order to reach the targeted $\alpha_2$,
it can be shown that if the voltage is kept constant, $N_2=(1 +
\epsilon)^{-\beta^{-1}}~N_1$, while if the number of applied pulses
is kept constant, the correction algorithm will be
$V_{pulse}^{(2)}$=(1 + $\epsilon$)~$V_{pulse}^{(1)}$ [with
$\epsilon= \frac{\partial
V_{pulse}}{\partial\alpha}~\frac{(\alpha_2-\alpha_1)}{V_{pulse}^{(1)}}$].
If we consider the data presented in Fig.4, in order to produce a
correction to the targeted $\alpha$ of 0.2 (from 0.1 to 0.3), as
-$\beta^{-1}$ $\simeq 14$ we obtain that $\epsilon$ $\simeq$ 0.16,
which indicates that the best strategy is to modify $V_{pulse}$ by a
16\% instead of increasing $\simeq$ 8 times the number of applied
pulses, which increases proportionally the time needed to correct
$\alpha$.

\begin{figure}
\vspace{7mm}
\centerline{\includegraphics[angle=0,scale=0.4]{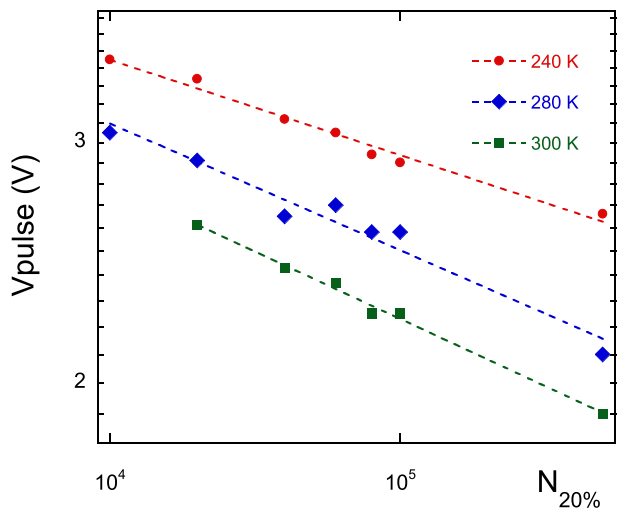}}
\vspace{-0mm}\caption{(Color online)V-N curves at different
temperatures for a failure criteria of 20 $\%$. Dotted lines are
fits corresponding to Eq.~\ref{eq:Basquin}. As for cyclic mechanical
stress experiments, lowering the temperature shifts the curves to
higher voltage stresses. } \vspace{-0mm} \label{fig:5}
\end{figure}

Other general result in material fatigue experiments, it is observed
that, due to the temperature dependence of plastic deformation, a
decrease in testing temperatures shifts the $S–N_\alpha$ curves
towards higher fatigue strengths.~\cite{Kohaut00} In
Fig.~\ref{fig:5} we have plotted our $V-N_\alpha$ curves considering
a fixed $\alpha = 20 \%$ where this behavior is also reproduced.
Similar results were obtained for other $\alpha_0$ values.

A possible interpretation of these results indicates that the
physics behind the electric field assisted propagation of vacancies
is similar to the propagation of defects produced during a cyclic
mechanical fatigue stress to a material. In fact if a fracture can
be considered as the consequence of an accumulation of interatomic
bonds ruptures our results are consistent with a framework were
oxygen diffuses producing correlated defects as twins or stacking
faults. As the remnant resistance of the metal-oxide interface is
considered proportional to the vacancy density near the
interface~\cite{Rozenberg10}, the observed increase of $\alpha$ with
the number of cycles is then a natural consequence of the oxygen
vacancy production rate.

\section{CONCLUSIONS}
We have studied the sensitivity of the remnant RS change to the
amplitude of cyclic voltage pulses at different temperatures and
number of pulses.

We showed that if an arbitrarily fixed percentage of resistance
change ($\alpha \geq \alpha_0$) is associated with the failure
criteria usually defined in mechanical tests, the electric field
equivalent stress-fatigue lifetime curves can be obtained for a
device. In this way, we provide the relation between the RS
amplitude and the number of applied pulses, at a fixed amplitude and
temperature. This relation can be used as the basis to built an
error correction scheme. Additionally, this similarity points out
that the evolution of the remnant resistance after a cyclic electric
field treatment related to the process of accumulation of vacancies
near the metal-oxide interface has a strong physical resemblance to
the propagation of defects in materials subjected to cyclic
mechanical stress tests.

\section{ACKNOWLEDGEMENTS}
We would like to acknowledge financial support by CONICET Grant PIP
112-200801-00930 and UBACyT 20020100100679 (2011-2014). We also
acknowledge V. Bekeris for a critical reading, and D. Gim\'enez, E.
P\'erez Wodtke and D. Rodr\'{\i}guez Melgarejo for their technical
assistance.


\end{document}